\documentclass[final,3p,times,twocolumn]{elsarticle}

\usepackage{graphics}

\usepackage{amssymb}

\journal{Journal of Luminescence}

\begin{document}

\begin{frontmatter}

\title{Spectroscopic investigations of a Ti:Tm:LiNbO$_3$ waveguide for photon-echo quantum memory}

\author[cal]{N. Sinclair\corref{cor1}\fnref{fn1}}
\ead{nsinclair@qis.ucalgary.ca}
\cortext[cor1]{Corresponding author}
\author[cal]{E. Saglamyurek\fnref{fn1}}
\fntext[fn1]{These authors contributed equally to this work.}
\author[pad]{M. George}
\author[pad]{R. Ricken}
\author[cal]{C. La Mela\fnref{fn2}}
\author[pad]{W. Sohler}
\author[cal]{W. Tittel}
\address[cal]{Institute for Quantum Information Science, and Department of Physics \& Astronomy, University of Calgary, Calgary, Alberta T2N 1N4, Canada}
\address[pad]{Angewandte Physik, Universit\"at Paderborn, 33098 Paderborn, Germany}
\fntext[fn2]{Current address: Department of Engineering Physics, \'{E}cole Polytechnique de Montr\'{e}al, Montr\'{e}al, Qu\'{e}bec H3C 3A7, Canada.}

\begin{abstract}
We report the fabrication and characterization of a Ti$^{4+}$:Tm$^{3+}$:LiNbO$_3$ optical waveguide in view of photon-echo quantum memory applications. In particular, we investigated room- and cryogenic-temperature properties via absorption, spectral hole burning, photon echo, and Stark spectroscopy. We found radiative lifetimes of 82 $\mu$s and 2.4 ms for the $^3$H$_4$ and $^3$F$_4$ levels, respectively, and a 44\% branching ratio from the $^3$H$_{4}$ to the $^3$F$_4$ level. We also measured an optical coherence time of 1.6 $\mu$s for the $^3$H$_6\leftrightarrow{}^3$H$_4$, 795 nm wavelength transition, and investigated the limitation of spectral diffusion to spectral hole burning. Upon application of magnetic fields of a few hundred Gauss, we observed persistent spectral holes with lifetimes up to seconds. Furthermore, we measured a linear Stark shift of 25 kHz$\cdot$cm/V. Our results are promising for integrated, electro-optical, waveguide quantum memory for photons.
\end{abstract}


\end{frontmatter}

\section{Introduction}
Quantum memory, i.e. the reversible transfer of quantum states between light and atoms, constitutes a key element for quantum repeaters \cite{Sangouard2009}, as well as linear optical quantum computing \cite{Kok2007}. Impressive experimental and theoretical progress has been reported over the past few years \cite{Hammerer2008,Lvovsky2009}, and gives hope that a workable quantum memory can eventually be built.

The \textit{photon-echo quantum memory} \cite{Tittel2009} is rooted in all-optical storage of classical data, investigated already thirty years ago \cite{Hesselink1979,Mossberg1982,Mitsunaga1991,Mitsunaga1992}. It relies on the interaction of light with a large ensemble of atoms with a suitably prepared, inhomogeneously broadened absorption line. In materials with natural broadening, this tailoring can be achieved by frequency-selectively removing absorbers from the ground to an auxiliary state using optical pumping techniques \cite{Pryde2000,Nilsson2004,Crozatier2004,Lauritzen2008}, possibly followed by controlled broadening of the resulting absorption lines, e.g. through position dependent Stark shifts.

The absorption of light in photon-echo quantum memory leads to mapping of its quantum state onto atomic coherence, which rapidly decays as the absorbers have different resonance frequencies. To recall the light, i.e. map the quantum state back onto light, the initial coherence has to be re-established. Two possibilities have been identified: In \emph{Controlled Reversible Inhomogeneous Broadening} (CRIB), the detuning of each absorber with respect to the light carrier frequency has to be inverted \cite{Moiseev2001,Nilsson2005,Alexander2006,Kraus2006}. Another possibility is to tailor the initial absorption line into an \emph{Atomic Frequency Comb} (AFC), resulting in re-emission of the stored light after a time that depends on the periodicity of the comb \cite{Hesselink1979,Afzelius2009,Riedmatten2008}. In both approaches, the memory efficiency can theoretically reach 100\% \cite{Moiseev2001,Kraus2006,Afzelius2009,Hetet2008}. For extended storage in CRIB or AFC (then also allowing for on-demand recall), the excited atomic coherence can be transferred temporarily to longer lived coherence, e.g. between hyperfine ground states. This reversible mapping of coherence can be achieved by means of two $\pi$-pulses \cite{Afzelius2009b}, or via a direct Raman transfer \cite{Hetet2008b,Moiseev2008,Gouet2009,Hosseini2009}.

Rare-earth-ion doped crystals (RE crystals) are promising material candidates for photon-echo quantum memory \cite{Tittel2009}. Key findings include the storage of qubit states encoded into attenuated laser pulses \cite{Riedmatten2008}, light storage with up to 66\% efficiency \cite{Hetet2008,Chaneliere2009,Amari2009,Hedges2009}, mapping of photonic quantum states onto collective atomic spin states \cite{Afzelius2009b}, and simultaneous storage of 64 photon modes \cite{Usmani2009}. Furthermore, observation of more than 30 s coherence time of a ground state hyperfine transition in Pr:Y$_2$SiO$_5$ \cite{Fraval2005} is highly promising for long-term storage in this crystal, as well as other RE crystals.

In this article we investigate a novel storage medium, a Ti$^{4+}$:Tm$^{3+}$:LiNbO$_3$ optical waveguide, in view of the requirements for photon-echo quantum memory. It combines interesting features arising from the specific RE dopant, the host material, as well as the waveguide structure.

\begin{figure}
\centering \includegraphics[width=0.8\columnwidth]{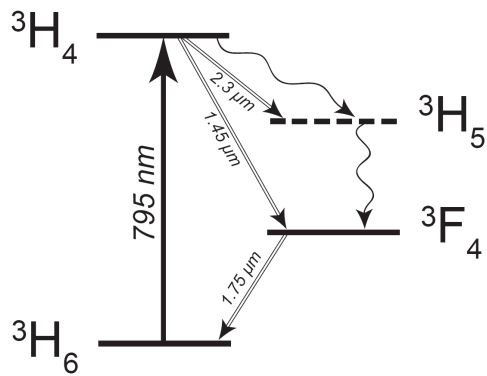}
\caption{Simplified energy level diagram of Tm:LiNbO$_3$ showing the electronic levels relevant to this work.}
\label{fig:tmlevels}
\end{figure}

As depicted in Fig. \ref{fig:tmlevels}, the thulium (Tm) $^3$H$_6\leftrightarrow{}^3$H$_4$ transition features absorption at $\sim$795 nm, which is a wavelength where air has minimal transmission loss, where photon pairs or entangled photon pairs are conveniently created \cite{Kwiat1995,Ribordy2001,Halder2009,Slater2009}, and where high-efficiency and simple-to-operate single photon detectors based on Silicon Avalanche Photodiodes are commercially available \cite{Si-APD}. This makes thulium-based quantum memory interesting for applications in quantum communication and linear optical quantum computing. Furthermore, the $^3$F$_4$ bottleneck state, whose energy is situated roughly half way between the $^3$H$_6$ and $^3$H$_4$ levels \cite{Gruber1989}, allows for efficient and broadband spectral tailoring \cite{Tian2001}. In addition, the application of suitably oriented magnetic fields results in long-lived nuclear spin levels forming $\Lambda$ systems, at least in the case of Tm:Y$_3$Al$_5$O$_{12}$ (Tm:YAG), and may be used for optical pumping or long-term storage \cite{Ohlsson2003,Guillot-Noel2005,Goldner2006,Seze2006,Louchet2007,Louchet2008}. Low-temperature spectroscopic investigations of Tm doped crystals have so far mostly concentrated on Tm:YAG (see also  \cite{Macfarlane1993}), and, very recently, Tm:LiNbO$_3$ \cite{Nunez1993-2,Mohan2007,Thiel2008,Thiel2009}. For Tm:YAG, they have already led to implementations of the AFC protocol \cite{Chaneliere2009,Chaneliere2009b}.

Due to lack of inversion symmetry, lithium niobate (LiNbO$_3$) crystals feature non-linear effects \cite{LNbnonlin}. The crystal symmetry also results in permanent electric dipole moments for RE states, i.e. the possibility to externally control resonance frequencies via Stark shifts \cite{Macfarlane2007,Hastings2006}, as required for CRIB. Low-temperature properties of RE doped LiNbO$_3$ have been characterized in  \cite{Mohan2007,Thiel2008,Hastings2006,Dierolf2001,Sun2002},  e.g. for the development of radio-frequency analyzers \cite{Colice2006}.

Given its non-linear properties, LiNbO$_3$ has become an important material for the telecommunication industry. Procedures to implement waveguides, either through proton exchange \cite{protonexchange}, or Titanium indiffusion \cite{Tiindiffusion}, have been developed, allowing for simple integration with fibre-optics components. In addition, given their small transverse dimensions of 5-10 $\mu$m (depending on the light wavelength), traveling wave electrodes can be spaced closely, resulting in commercial intensity and phase modulators with switching ($\pi$) voltages of only a few volts, and switching times below 100 picoseconds \cite{LNbtelecom}. Furthermore, as light intensities inside these waveguides can be very large, strong non-linear interactions are possible, including frequency up-conversion of photons \cite{Tanzilli2005}. For photon-echo quantum memory, these properties promise simple integration with fibre quantum networks, sub-ns Stark shifting, and large Rabi frequencies, which will benefit optical pumping procedures in all photon-echo quantum memory protocols. LiNbO$_3$ waveguides have been used for studies relevant to quantum memory in \cite{Hastings2006,Staudt2007,Staudt2007b,Delfan2009}.

The remainder of this article is structured as follows: In Sec. \ref{waveguide fabrication}, we describe and characterize the fabrication process of the waveguide. Next, in Sec. \ref{absorption_profiles}, we discuss polarization dependent absorption profiles measured at room and cryogenic temperatures. After a description of the experimental setup used for all following low-temperature measurements in Sec. \ref{experimental_setup}, Sec. \ref{population} focuses on relevant radiative lifetimes and branching ratios. Our investigations of optical coherence times, and the limitation of spectral hole burning imposed by spectral diffusion are presented in Sec. \ref{coherence}, followed by Stark-effect-based modification of narrow absorption lines. Finally, in Sec. \ref{discussion}, we discuss our results in view of photon-echo quantum memory. This concludes the article.

\section{Waveguide fabrication and characterization}\label{waveguide fabrication}
\emph{Tm-diffusion doping of LiNbO$_{3}$}--
Commercially available 0.5 mm thick Z-cut wafers of undoped optical grade congruent lithium niobate (CLN) were used as starting material. Samples of 12 mm x 30 mm size were cut from these wafers and doped by thulium near the +Z-surface before waveguide fabrication. The doping was achieved by in-diffusing a vacuum deposited (electron-beam evaporated) Tm layer of 19.6 nm thickness. The diffusion was performed at 1130$^\circ$C during 150 h in an argon-atmosphere followed by a post treatment in oxygen (1 h) to get a full re-oxidization of the crystal.

\begin{figure}
\centering \includegraphics[width=0.95\columnwidth]{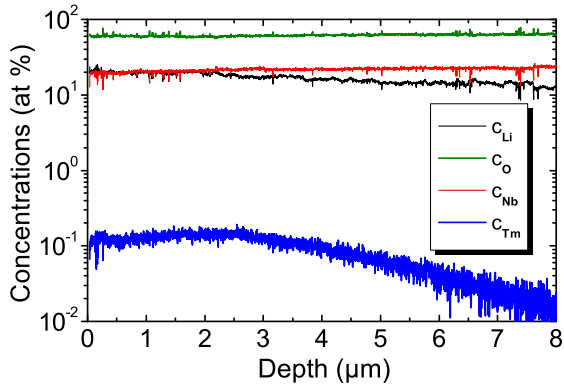}
\caption{Measured concentrations of Tm, Li, Nb, and O versus depth, using SNMS with 700 eV Ar-ions.}
\label{fig:concentrations1}
\end{figure}

To determine the diffusion coefficient of Tm into Z-cut CLN, secondary neutral mass spectroscopy (SNMS) was performed using 700 eV Argon-ions for ion milling. Ions and electrons were extracted from the plasma source with a duty cycle of 4:1 at a rate of 320 kHz to avoid charging of the insulating CLN-substrate. SNMS was chosen instead of secondary ion mass spectroscopy (SIMS) to significantly reduce matrix effects (see e.g. \cite{Bubert2002}). In Fig. \ref{fig:concentrations1}, the concentration profiles versus depth have been recorded for thulium (Tm), lithium (Li), niobium (Nb) and oxygen (O). Interestingly, the Li-concentration slightly increases towards the surface, although it is expected that Tm occupies regular Li-sites similar to Er-ions when incorporated in CLN by diffusion \cite{Quintanilla2008}.

\begin{figure}
\centering \includegraphics[width=0.95\columnwidth]{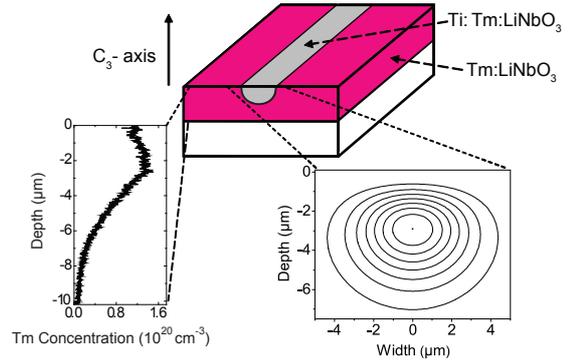}
\caption{Scheme of the waveguide geometry with the measured Tm concentration profile on the left and the calculated intensity distribution of the fundamental TM-mode at 795 nm wavelength on the right. Iso-intensity lines are plotted corresponding to 100\%, 87.5\%, 75\% etc. of the maximum intensity.}
\label{fig:wguide}
\end{figure}

In Fig. \ref{fig:wguide}, the Tm concentration is plotted on a linear scale versus the depth. The slight dip close to the surface is unexpected and needs further investigations. Fitting a Gaussian profile to the concentration curve leads to a 1/e-penetration depth $\textrm{d}_{1/\textrm{e}}$ of about 6.5 $\mu $m. Together with the diffusion parameters, a diffusion coefficient $\textrm{D}=\textrm{d}_{1/\textrm{e}}^2/(4\textrm{t})=0.07 \mu$m$^2/\textrm{h}$, where t denotes the diffusion time, was evaluated for 1130$^\circ$C. At this temperature, this is seven times larger than the corresponding coefficient for Erbium-diffusion into CLN \cite{Baumann1997}. The maximum Tm concentration of about 1.35$\cdot 10^{20}$ cm$^{-3}$ corresponds to a concentration 0.74 mole \%, which - according to Ref. \cite{Quintanilla2008} - is considerably below the solid solubility of Tm in CLN.


\emph{Ti-indiffused waveguides in Tm:LiNbO$_3$}--
On the Tm-diffusion doped surface of the substrate, a 40 nm thick titanium (Ti) layer was deposited using electron-beam evaporation. From this layer, 3.0 $\mu$m wide Ti stripes were photolithographically defined and subsequently in-diffused at 1060$^\circ$C for 5 h to form 30 mm long optical strip waveguides. In the wavelength range around 775 nm, the waveguides are single mode for TE- and TM-polarization (see Fig. \ref{fig:wguide}).

\begin{figure}
\centering \includegraphics[width=0.95\columnwidth]{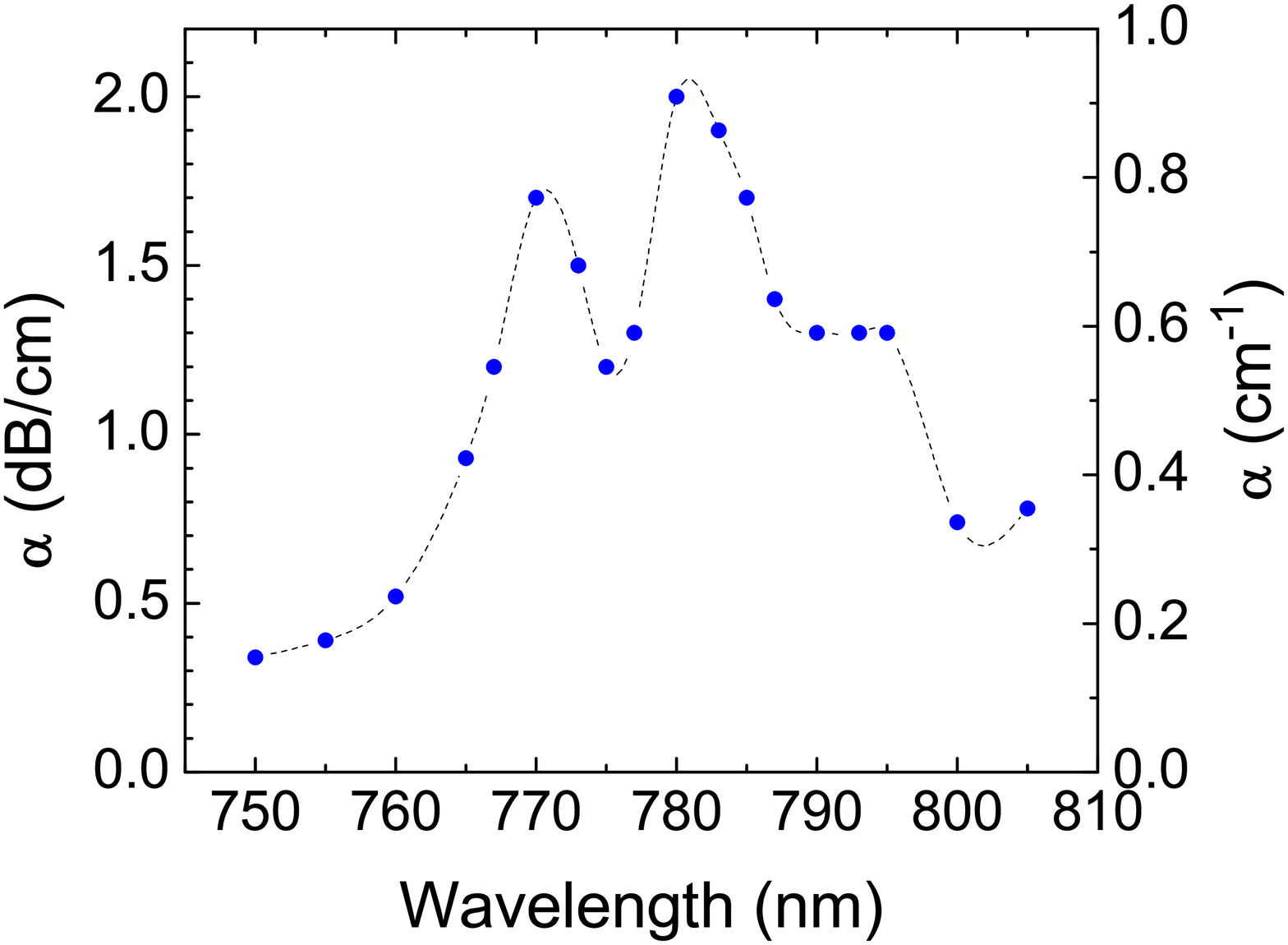}
\caption{Measured loss coefficient ($\alpha$) for TM-polarization as a function of wavelength. Data points are connected by spline fitting as a guide for the eye.}
\label{fig:absorptioncoefficients1}
\end{figure}

The waveguide end faces were carefully polished normal to the waveguide axis, forming a low-finesse resonator. This allows to determine the total waveguide propagation loss at room temperature, including absorption and scattering loss,  by the Fabry-Perot method \cite{Regener1985}. A stabilized, single frequency Ti:Sapphire laser was used to measure the transmission of the low-finesse waveguide resonator as function of a small temperature change at a number of fixed wavelengths in the range between 750 nm and 807 nm. From the contrast of the measured Fabry-Perot response, the propagation loss was deduced for all wavelengths \cite{Regener1985}. The results are presented in Fig. \ref{fig:absorptioncoefficients1} for TM-polarization.

In addition, the waveguide propagation loss was determined at room temperature and 729 nm wavelength, where negligible absorption by the Tm-ions can be expected. Therefore, the measured loss coefficients reflect the scattering loss alone; they are 0.2 dB/cm for TE- as well as for TM-polarization. As the scattering loss is only weakly dependent on the wavelength, it can be regarded as a background for the Tm-induced absorption loss.

We repeated the off-resonant, polarization dependent loss measurement at 3.5 K, leading to strongly polarization dependent transmission. The exact origin of this difference compared to the room-temperature measurement requires further investigation. For this as well as all subsequent low-temperature measurements, the sample was cut to 15.7 mm and repolished. All room temperature measurements have been performed using a 30 mm long waveguide.

\section{Spectroscopy of inhomogeneous broadening}
\label{absorption_profiles}

Reversible atom-light interaction requires absorption on the transition between the lowest lying Stark levels in the ground and excited electronic states. To gain information about $^3$H$_4$ and $^3$H$_6$ Stark splittings, we injected weak, broadband, polarized light into the waveguide. Using wave plates, we set the polarization to TE or TM, and measured power spectra of transmitted light using an optical spectrum analyzer.

\begin{figure}
\centering \includegraphics[width=0.95\columnwidth]{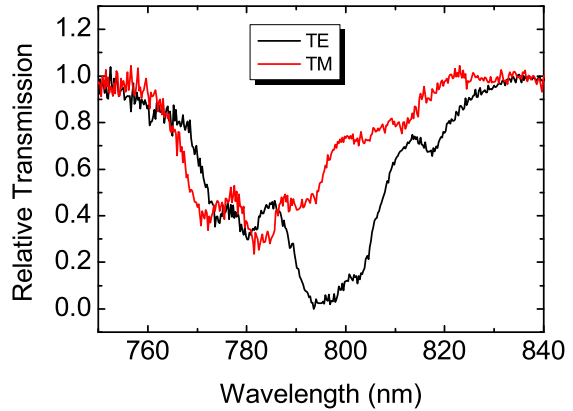}
\caption{(Color online) Relative transmission through the Ti:Tm:LiNbO$_3$ waveguide for TM- and TE-polarization, respectively, as a function of wavelength. The resolution bandwidth of the optical spectrum analyzer used in this experiment was 2 nm due to the low spectral power density of the thermal radiator.}
\label{fig:absorptioncoefficients2}
\end{figure}

\emph{Room temperature characterization}--
The transmission through a 30 mm long waveguide for TE and TM-polarized light is shown in Fig. \ref{fig:absorptioncoefficients2}. It has been normalized to the incident spectral power density of the broadband tungsten lamp used in this experiment. We observe broad absorption, reflecting different transitions between Stark levels in the $^3$H$_6$ and $^3$H$_4$ multiplets (superimposed with inhomogeneous broadening), and the thermal distribution of the population in the electronic ground state. In addition, a strong polarization dependence of absorption is observed, confirming previous studies performed on bulk crystals \cite{Nunez1993-2,Cantelar2008}.

\begin{figure}
\centering \includegraphics[width=0.95\columnwidth]{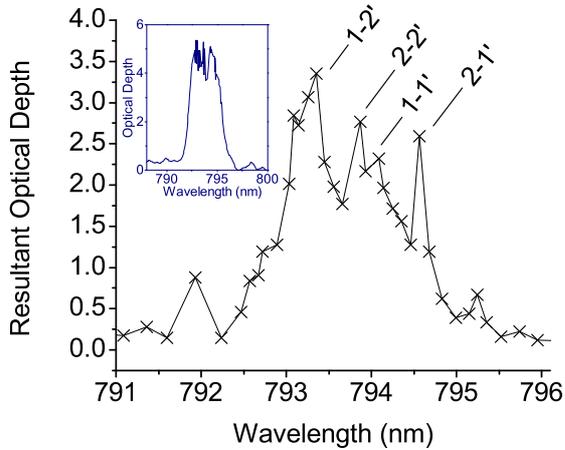}
\caption{Main figure: Absorption profile (after partial bleaching) at 3.5 K obtained using a single frequency laser. Indicated are transitions between different Stark levels: 1 and 2 denotes the lowest energy levels within the $^3$H$_6$ multiplet, primed labels represent the lowest $^3$H$_4$ levels. Inset: The same inhomogeneous broadening probed using weak broadband light.}
\label{fig:inh}
\end{figure}

\emph{Low temperature characterization}-
A similar measurement at 3.5 K resulted, for close-to TM polarization, in the absorption profile shown in the inset of Fig. \ref{fig:inh}. Due to the large optical depth of our sample, reflecting redistribution of population in the ground state Stark levels, we observe an almost flat-lined spectrum.

To resolve the Stark splittings, coherent laser light was coupled into the waveguide, and its frequency was swept between 791 and 796 nm in 0.1 nm steps. The laser intensity was optimized to resolve different Stark transitions after having partially bleached the absorption line. The resulting optical depth, shown in the main plot of Fig. \ref{fig:inh}, was determined at each measured wavelength upon normalization to the probe light. Using results from \cite{Thiel2009,Johnston1969}, we can identify four transitions between the ground and excited state Stark multiplets, with splittings of 0.48 $\pm$ 0.05 nm and 0.93 $\pm$ 0.05 nm, respectively (i.e. 7.6 $\pm$ 0.8 cm$^{-1}$ and 14.7 $\pm$ 0.8 cm$^{-1}$, respectively). This indicates the presence of a zero-phonon line in the high wavelength region. According to calculations taking into account the observed splittings, only a small fraction of atomic population $\sim$1\% occupies excited states in the ground Stark multiplet at $\sim$ 3 K.

\section{Narrow-band spectroscopic investigations}
\subsection{Experimental setup}
\label{experimental_setup}
\begin{figure}
\centering \includegraphics[width=0.95\columnwidth]{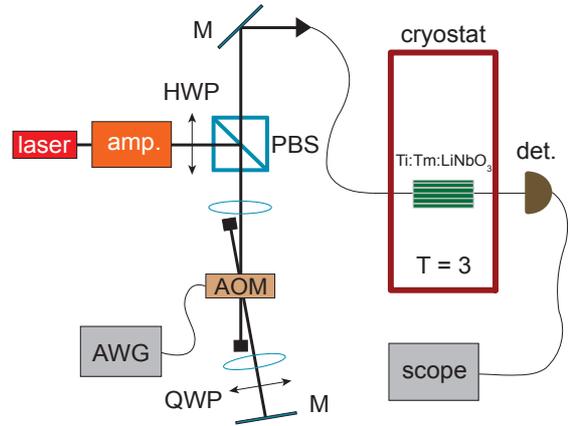}
\caption{Schematics of the experimental setup used for the narrow-band measurements at cryogenic temperature. HWP: half-wave plate; QWP: quarter-wave plate; PBS: polarization beam splitter; AOM: acousto-optic modulator; AWG: arbitrary waveform generator; M: mirror; amp.: optical amplifier; det.: detector; scope: oscilloscope.}
\label{fig:setup}
\end{figure}

A schematic of the experimental setup used for the low-temperature spectroscopic measurements described hereafter is depicted in Fig. \ref{fig:setup}.

A continuous wave, external cavity diode laser was tuned to 795.520 nm wavelength, where we found optical coherence properties of our sample in the absence of a magnetic field to be promising. The laser's linearly polarized output was amplified, and frequency and intensity modulated using a 400 MHz acousto-optic modulator in double-pass configuration, driven by a 10 GS/s arbitrary waveform generator and an amplifier. This allowed creating optical pulses with peak powers up to $\sim$4 mW and durations between 20 ns and 500 ms for photon-echo sequences and spectral hole burning. Here and henceforth, all pulse powers are specified at the input of the cryostat. After passing a $\lambda/2$ wave plate, the light was coupled into a single-mode optical fiber, and sent into the 3.5 $\mu$m wide, single mode Ti:Tm:LiNbO$_3$ waveguide. The light's polarization, which could be partially controlled using the wave plate, was set as to maximize the transmission, i.e. as to minimize the distance to TM. To inject and retrieve the light from the waveguide, optical fibres were butt-coupled against its front and end face using high precision translation stages, resulting in fibre-to-fibre coupling loss around 15 dB. The waveguide was placed in a pulse tube cooler and cooled to 3 K. A superconducting coil allowed generating magnetic fields along the crystal's C$_3$-axis, and electric fields could be generated along the same direction using aluminum electrodes that were firmly pressed against the crystal. Transmitted pulses and echoes were detected using either a sensitive 10 MHz, or a fast 125 MHz photodetector, which was then connected to a 3 GHz bandwidth oscilloscope with 10 GS/s sampling rate. All measurements conducted at zero magnetic field were repeated every 100 ms, and generally averaged 300 times. Upon magnetic field application, the repetition period was set to 60 s, which was required to avoid accumulation effects, and the number of averages was reduced to 25.

\subsection{Population relaxation dynamics}\label{population}

To optimize optical pumping strategies, and determine possibilities for long-term storage, it is important to examine the relaxation avenues and dynamics of population in atomic levels involved in the light-atom interaction. The level scheme for thulium doped crystals in the case of zero magnetic field is depicted in Fig. \ref{fig:tmlevels}. Upon excitation of the lowest lying Stark level within the $^3$H$_4$ manifold, atoms will eventually decay back to the ground state, either directly, or via the $^3$H$_5$ and $^3$F$_4$ levels. As the radiative lifetime of the $^3$H$_5$ level is short compared to the lifetimes of the $^3$H$_4$ and $^3$F$_4$ levels \cite{Nunez1993-2}, we model atomic decay, i.e. radiative lifetimes and branching ratio (given by the rate of decay from $^3$H$_4$ to $^3$F$_4$ relative to the overall decay from $^3$H$_4$), using a simplified three-level scheme comprising only the $^3$H$_6$ ground state, the $^3$F$_4$ bottleneck state, and the $^3$H$_4$ excited state.

\begin{figure}
\centering \includegraphics[width=0.95\columnwidth]{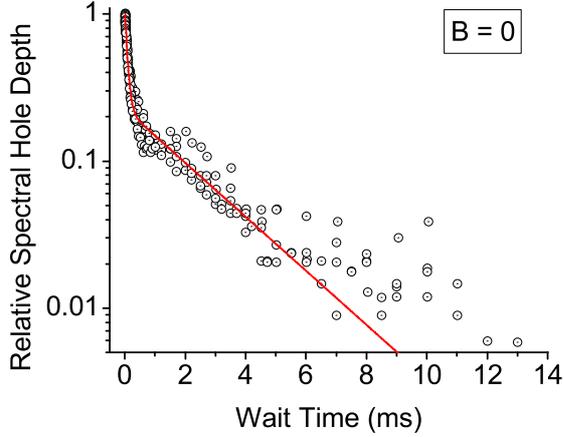}
\caption{Spectral hole decay under zero magnetic field. Plotted circles denote the normalized spectral hole depth as a function of the waiting time between burning and reading pulses. Two exponential decays are easily identified, yielding radiative lifetimes of 82 $\mu$s and 2.4 ms for the $^3$H$_4$ and $^3$F$_4$ levels, respectively. The branching ratio into the $^3$F$_4$ level is approximately 44\%.}
\label{fig:shbt1}
\end{figure}

As material absorption is dependent upon population differences between the states coupled by the probe light, we can assess population dynamics through time-resolved spectral hole burning. To this end, we first transfer population to the excited state using a 5-10 $\mu$s long \textit{burning} pulse with peak power of 7 $\mu$W, and then probed the shape and depth of the created spectral hole after a \textit{waiting time} ranging from 10 $\mu$s to 15 ms using a chirped \textit{reading} pulse. The power of the reading pulse, around 1 $\mu$W, was chosen as to not alter the population distribution created by the burning pulse. Fig. \ref{fig:shbt1} depicts the time dependent depth of the spectral hole, which we found to be proportional to the hole area (i.e. the effect of spectral diffusion, leading to a waiting-time-dependent hole width, was not visible in this measurement).

We model the decay using three-level rate equations, leading to

\begin{equation}
\frac{\Delta \textrm{d} (\textrm{t})}{\Delta \textrm{d}(\textrm{0})} = (1-\textrm{B})\textrm{e}^{-\textrm{t}/\textrm{T}_{1e}}+\textrm{Be}^{-\textrm{t}/\textrm{T}_{1b}}
\label{threeleveldecay}
\end{equation}

\noindent where $\Delta\textrm{d}(\textrm{t})$ denotes the reduction of optical depth d = $\alpha$L at the centre of the spectral hole at time t after burning, $\alpha$ is the absorption coefficient, L the sample length, T$_{1e}$ and T$_{1b}$ are the T$_1$ lifetimes of the excited and bottleneck states, respectively, and B $=\frac{\beta}{2}\frac{\textrm{T}_b}{\textrm{T}_b-\textrm{T}_e}$ identifies the branching ratio $\beta$. We find the time-dependent population difference to be characterized by a fast exponential decay with a lifetime of T$_{1e}$ = 82 $\pm$ 2 $\mu$s, and a slower exponential decay from $^3$F$_4$, characterized by T$_{1b}$ = 2.364 $\pm$ 0.198 ms. Both decay constants agree with previous measurements \cite{Cantelar2005}. Furthermore, we find a branching ratio of 0.436 $\pm$ 0.017.

\begin{figure}
\centering \includegraphics[width=0.95\columnwidth]{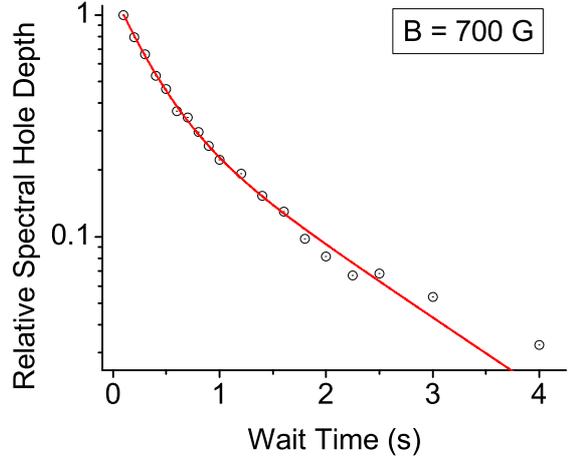}
\caption{Spectral hole depth as a function of the waiting time for a magnetic field of 700 G. The observation of two exponential decays indicates the existence of long-lived, ground state sub-levels.}
\label{fig:shbhf}
\end{figure}

\begin{figure}
\centering \includegraphics[width=0.95\columnwidth]{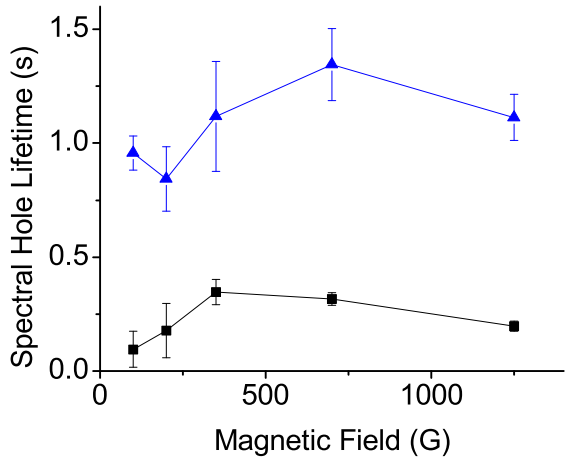}
\caption{Magnetic field dependence of the two decay times extracted from measurements of the spectral hole depth as a function of the waiting time. The case B = 700 G is shown in Fig. \ref{fig:shbhf}.}
\label{fig:t1bfield}
\end{figure}

We also performed hole burning studies under application of magnetic fields between 100 and 1250 Gauss, oriented parallel to the crystal C$_3$-axis. For these measurements, we increased the burning time to 500 ms, and varied the waiting time between 100 ms and 6 s so that population in the $^3$H$_4$ and $^3$F$_4$ levels could be ignored. As depicted in Fig. \ref{fig:shbhf}, this allowed the observation of spectral holes persisting during waiting times of up to seconds, with the decay of the hole depth being again well described by the sum of two exponentials. Furthermore, the two decay times change with magnetic field, as depicted in Fig. \ref{fig:t1bfield}. This indicates the appearance of two, magnetic field dependent atomic levels with long lifetimes, and suggests maximum lifetimes at magnetic fields around 600 G.

While we have not been able to identify the levels involved in the long-term storage, e.g. through the observation of additional holes or anti-holes that indicate field-induced level splitting \cite{Macfarlane1988}, we attribute this observation to storage of population in nuclear hyperfine levels, and levels arising through super-hyperfine interaction with neighboring Lithium or Niobium ions \cite{Thiel2008,Thiel2009}. We point out that the direct relaxation between different nuclear spin states in the excited and ground state multiplets is likely to be forbidden in the case where the magnetic field is parallel to the crystal C$_3$-axis \cite{Guillot-Noel2005}. The observed long-lived storage thus probably involves a (spin mixing) relaxation pathway including the $^3$H$_5$ level.

Furthermore, relaxation between different ground states, which gives rise to the curves shown in Figs. \ref{fig:shbhf} and \ref{fig:t1bfield}, is likely to involve several contributions, including spin-lattice relaxation, spin-spin flip flops between neighboring Tm ions, or interactions of Tm ions with other magnetic impurities \cite{Davids1964,Larson1966,Kurkin1980, Bottger2006,Hastings2008a,Hastings2008b}. Further studies at different temperatures and with crystals with smaller Tm ion concentration are in progress.

To summarize this section, we have observed a short lifetime of the $^3$H$_6$ level compared to the $^3$F$_4$, and the $^3$H$_4$ sub-levels. We have also identified a large branching ratio from $^3$H$_4$ to $^3$F$_4$. These observations identify two possible approaches to spectral tailoring, either involving the bottleneck state, or the magnetic field dependent ground states as an auxiliary state.
In this context, the existence of the bottleneck state is of particular interest. Indeed, due to the large energy gap relative to the $^3$H$_6$ and $^3$H$_4$ electronic levels, population pumped into this state during spectral tailoring does not interact with light coupling the ground and excited states. This allows tailoring of the initial absorption line over a spectral interval that is in principle only limited by the inhomogeneous broadening of the $^3$H$_6\leftrightarrow{}^3$H$_4$ transition, and not by energy spacings in the ground or excited levels \cite{Nilsson2004}. In other words, use of the bottleneck state for optical pumping could allow storage of short pulses of light with spectral width exceeding GHz.

Similarly, our observations suggest possibilities to reversibly map optically excited coherence (on the $^3$H$_6\leftrightarrow{}^3$H$_4$ transition) onto $^3$H$_6\leftrightarrow{}^3$F$_4$ coherence, or long-lived ground state coherence.

\subsection{Optical coherence and spectral diffusion}\label{coherence}
For photon-echo quantum memory, the initial inhomogeneous absorption line must be tailored into one or more narrow lines using frequency selective optical pumping. In the case of CRIB, the width of the resulting spectral feature determines the time quantum information can be stored in optical coherence \cite{Sangouard2007}. For AFC, it determines the spacing of the teeth in the comb structure, which, in turn, sets the storage time in optical coherence \cite{Afzelius2009}. Material dependent constraints to this time arise from non-zero homogeneous line width $\Gamma_{hom}$, which is limited by natural broadening: $\Gamma_{hom}^{nat} = 1/(2\pi\textrm{T}_{1e})\approx 1.9$ kHz, phonon broadening \cite{Meltzer2005}, as well as long-term spectral diffusion \cite{Bottger2006,Mims1961,Bai1989}.

\begin{figure}
\centering \includegraphics[width=0.95\columnwidth]{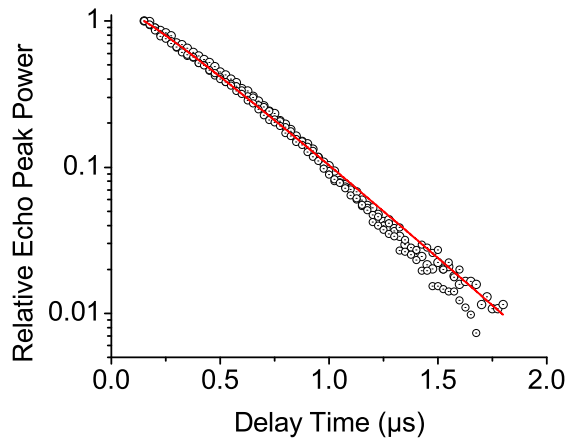}
\caption{Two pulse photon-echo peak powers measured under zero magnetic field. Plotted circles are normalized echo powers as a function of the delay time between the two pulses. Fitted is the Mims expression giving a coherence time of 1.6 $\mu$ with nearly absent spectral diffusion.}
\label{fig:2ppe}
\end{figure}

To assess the short-term homogeneous line width, we employed two-pulse photon-echoes. Two 20 ns long pulses, with peak powers $\sim 3$ mW that maximized the observed echo, were sent into the thulium waveguide, and the relative \textit{delay} was varied from 100 ns to 1.8 $\mu$s in steps of 25 ns. Fig. \ref{fig:2ppe} depicts the resulting peak echo powers for the case of zero magnetic field. We fit the decay of the peak echo intensity I (which is proportional to power) with the Mims expression \cite{Mims1968}:
\begin{equation}
\textrm{I}=\textrm{I}_{0}\textrm{exp}(-4\textrm{t}/\textrm{T}_{2})^x
\label{Mims}
\end{equation}
\noindent
where $\textrm{I}_{0}$ denotes the maximum echo intensity, T$_2$ is the phase memory (coherence) time and $x$ characterizes spectral diffusion. The fit revealed a phase memory time of 1.580 $\pm$ 0.008 $\mu$s, equivalent to a homogeneous line width of $\sim$200 kHz, and a spectral diffusion parameter $x$ of 1.072 $\pm$ 0.009. We obtained similar results for non-zero fields up to 250 Gauss. This indicates that the short-term homogeneous line width at 3 K is dominated by phonon scattering.

Beyond the short-term homogeneous line width, the narrowest spectral feature that can be generated through optical pumping is limited by spectral diffusion. Spectral diffusion is caused by fluctuations of each ion's transition frequency due to the dynamic nature of the ion's environment. This causes broadening of the homogeneous line width over time. The three pulse photon-echo (3PPE), or stimulated photon-echo, is a useful tool to investigate spectral diffusion \cite{Bottger2006}.

\begin{figure}
\centering \includegraphics[width=0.95\columnwidth]{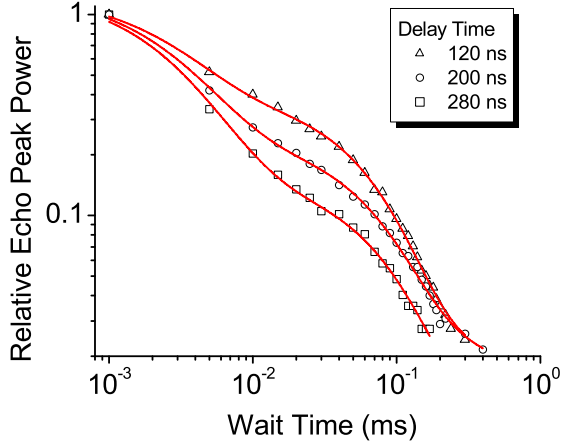}
\caption{Decay of stimulated echo with waiting time for delay times of 120 ns (triangle), 200 ns (circle) and 280 ns (square), respectively.}
\label{fig:stimecho}
\end{figure}

We performed a series of 3PPE experiments to probe spectral diffusion in our sample. All experiments were carried out at zero magnetic field. In the measurements, for three different \textit{delay} settings between the first two pulses, we varied the \textit{waiting} time, i.e. the time between the second and the third pulse, from 1 $\mu$s to 400 $\mu$s with 5-10 $\mu$s increments. The echo peak power was measured for each delay and waiting time, and each set of measurements (i.e. measurements with a specific delay time) was normalized to the echo peak power at 1 $\mu$s waiting time. The results are illustrated in Fig. \ref{fig:stimecho}.

To interpret the data, we employed the spectral diffusion model discussed in Ref. \cite{Bottger2006}. In this model, the peak intensity I of the stimulated echo is determined by the relative dephasing during the delay time $\textrm{t}_{D}$, the decay of the excited level population during the waiting time $\textrm{t}_W$, and diffusion mechanisms which broaden the line into the time-dependent effective line width $\Gamma_{eff}$. In its general form, the 3PPE intensity can be written as:
\begin{equation}
\textrm{I}(\textrm{t}_W,\textrm{t}_D)=\textrm{I}_{0}\textrm{F}(\textrm{t}_{{W}})\textrm{exp}\big\{-4\pi\textrm{t}_{D}\Gamma_{eff}(\textrm{t}_{D},\textrm{t}_W)\big\}
\label{spectral diffusion}
\end{equation}
\noindent
where $\textrm{I}_{0}$ denotes the maximum echo intensity,  $\textrm{F}(\textrm{t}_W)=(1-\textrm{B})\textrm{e}^{-2\textrm{t}_W/\textrm{T}_{1e}}+\textrm{Be}^{-2\textrm{t}_W/\textrm{T}_{1b}}$ describes the population decay during $\textrm{t}_W$, and B is defined as in Eq. \ref{threeleveldecay}. The effective line width describes spectral diffusion during both the delay and waiting time, and is given by:

\begin{equation}
\Gamma_{eff}(\textrm{t}_{D},\textrm{t}_W) =\\ \Gamma_{0}+\frac{1}{2}\Gamma_{SD}\big[\textrm{R}\textrm{t}_{D}+(1-\textrm{exp}(-\textrm{R}\textrm{t}_W))\big]
\label{effective linewidth}
\end{equation}
\noindent
where $\Gamma_{0}$ is the short-term line width in absence of spectral diffusion, and $\Gamma_{SD}$ denotes the maximum additional line width due to spectral diffusion, which occurs at rate R.

To fit our data, we fixed the bottleneck level lifetime to 2.4 ms, obtained through the spectral hole burning measurements discussed above.
This was required due to the echo intensity reaching noise level after 200 $\mu$s, which is short for the fit to generate a reliable lifetime. We also used the data obtained from the 2PPE measurements to extract an intrinsic homogeneous line width of $\Gamma_{0} = $152 $\pm$ 2 kHz via Eq. \ref{spectral diffusion}, in reasonable agreement with the result of $\sim$ 200 kHz found via Eq. \ref{Mims}.
Fitting all delay settings yielded an average excited level lifetime and branching ratio of 83 $\pm$ 8 $\mu$s and 0.23 $\pm$ 0.03, respectively, in reasonably good agreement with the more reliable values obtained from the spectral hole burning measurements.

Furthermore, the fit yielded $\Gamma_{SD}=930\pm 51$ kHz and a spectral diffusion rate of 227 $\pm$ 24 kHz, and the diffusion model thus predicts that the effective line width 
\begin{equation}
\Gamma_{eff}(\textrm{t}_{W})=\Gamma_{0} + \frac{\Gamma_{SD}}{2}\big[1-\textrm{exp}\{-\textrm{Rt}_{W}\}\big]
\label{saturationlinewidth}
\end{equation}
\noindent
saturates at around 630 kHz after a waiting time of $\sim$50 $\mu$s. When assessed through spectral hole burning, this leads to a homogeneous line width of $\Gamma_0+\Gamma_{SD}$=1082 kHz \cite{Staudt2006}.

To verify this prediction, we performed another series of spectral hole burning measurements with burning and waiting times of 5 $\mu$s and 50 $\mu$s, respectively. Varying the power of the burning pulse from 400 to 4 $\mu$W, and extrapolating the hole width to zero burning power \cite{Maniloff1995}, we find a homogeneous line width of 1.5 $\pm$ 0.1 MHz. Taking into account laser frequency jitter of $\sim1$ MHz, this is consistent with the prediction from the spectral diffusion model.

Summarizing these results, we have identified spectral diffusion as the limiting factor for Ti:Tm:LiNbO$_3$ to storage of quantum information in optical coherence. For instance, assuming an AFC with 3 MHz teeth spacing, the storage time in optical coherence would be limited to $\sim$ 300 ns. We expect that the application of a magnetic field and the decrease of temperature will lead to an improvement of the short-term line-width along with a reduction of spectral diffusion, similar to what has been observed for Tm:LiNbO$_3$ bulk crystals \cite{Thiel2009}.

\subsection{Stark effect}\label{stark}
Electric field control can constitute a key ingredient in photon-echo quantum memory and quantum information processing. It enables controlled manipulation of resonance frequencies of absorbers for storage and recall, e.g. for controlled reversible inhomogeneous broadening of narrow absorption lines in CRIB, and quantum state engineering such as pulse compression/decompression \cite{Moiseev2009b}. This control is governed by the interaction of applied external fields with permanent electric dipole moments. Provided the dipole moment is different for the ground and excited state, a shift in the resonance frequency $\Delta\omega$ occurs: 

\begin{equation}
\centering \Delta\omega = \frac{\chi}{\hbar}\Delta\overrightarrow{\mu_{e}}\cdot\overrightarrow{E}
\label{starkeqn}
\end{equation}
\noindent
where $\Delta\overrightarrow{\mu_{e}}$ denotes the difference in electric dipole moments for the states connected by the probe light, $\overrightarrow{E}$ is the applied electric field, and $\chi$ is the Lorentz correction factor. This is also known as the DC Stark effect \cite{Macfarlane2007}. 

To observe the frequency shift for different electric fields, we first burned a spectral hole, then applied a variable voltage parallel to the crystal $C_3$ axis, and assessed the displacement of the hole using a weak, chirped, read pulse, as detailed in \cite{Hastings2006}.

\begin{figure}
\centering \includegraphics[width=0.95\columnwidth]{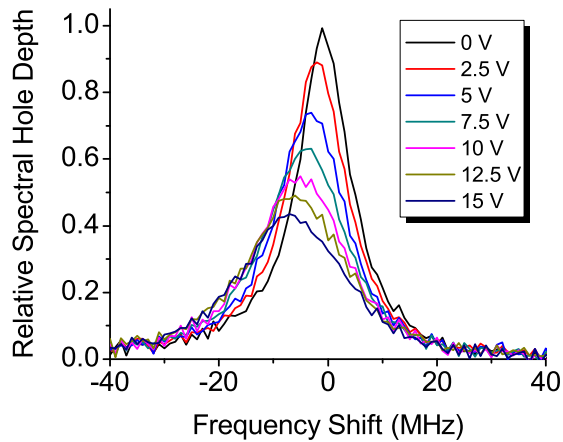}
\caption{Change of a spectral hole under application of different voltages. We attribute the broadening of the spectral hole with increased voltage to the large inhomogeneity of the electric field at the beginning and end of the LiNbO$_{3}$ waveguide.}
\label{fig:starkhole}
\end{figure}

\begin{figure}
\centering \includegraphics[width=0.95\columnwidth]{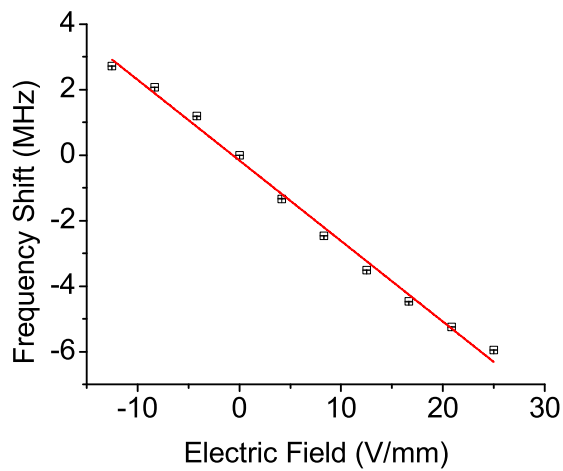}
\caption{Shift of transition frequency of the center of a spectral hole as a function of applied electric field, yielding a shift of 24.6 $\pm$ 0.7 kHz$\cdot$cm/V.}
\label{fig:starkdipole}
\end{figure}

As shown in Figs. \ref{fig:starkhole} and \ref{fig:starkdipole}, we observe a linear frequency shift of 24.6 $\pm$ 0.7 kHz$\cdot$cm/V. For example, an electric field of 100 V/mm leads to a displacement of the resonance frequency by 25 MHz. In the case of a Tm waveguide, where electrodes can be spaced as closely as 10 $\mu$m, this requires the application of 1 Volt. Since low voltages can be switched rapidly with ease, waveguides provide the ability to reversibly broaden and manipulate absorbers within hundreds of picoseconds.

\section{Discussion and conclusion}\label{discussion}
\label{conclusion}

To conclude, our findings demonstrate the suitability of Ti:Tm:LiNb$_3$ waveguides cooled to 3 K for implementations of photon-echo quantum memory protocols. Level structure, lifetimes, and branching ratios allow tailoring of the natural, inhomogeneously broadened absorption profile -- either via optical pumping into the $^3$F$_4$ bottleneck state (then possibly allowing storage of nanosecond pulses), or into one of the long-lived ground states that appear under the application of magnetic fields. The minimum width of spectral holes of around one MHz, as determined by spectral diffusion, will limit storage of quantum information in optical coherence to a few hundred nanoseconds. While longer times may be achievable at lower temperature, this is still sufficient for mapping coherence onto long-lived ground state coherence, as Rabi frequencies exceeding hundred MHz can be obtained, due to the high power densities achievable inside waveguiding structures. We point out that ground state coherence of 300 $\mu$s have been reported for Tm:YAG \cite{Louchet2008}, but investigations for Ti:Tm:LiNbO$_3$ remain to be done. Finally, the existance of a linear Stark shift, together with the possibility to space electrodes closely, allows shifting of resonance frequencies by more than 100 MHz within sub-nanosecond times, thus enabling novel phase control techniques.

Interestingly, the lifetimes and branching ratio found in our study differ from those reported for Tm:LiNbO$_{3}$ bulk crystals probed at 794.22 nm wavelength and 1.7 K \cite{Thiel2008,Thiel2009}. The difference could be due to the addition of Titanium to our sample, which may alter radiative or non-radiative decay channels \cite{Dierolf2001}, to wavelength-dependent spectroscopic properties as suggested in \cite{Thiel2009}, or to concentration dependence \cite{Quintanilla2008b}. 

Finally, we point out that RE doped waveguides, here Ti:Tm:LiNbO$_3$, are not only interesting for quantum state storage using a photon-echo approach, but also for other approaches, e.g. based on electromagnetically induced transparency and slow light \cite{Lvovsky2009}.

\section{Acknowledgements}
WT gratefully remembers discussions with Drs. Olivier Gouillot-No\"{e}l, Ivan Lorger\'{e}, and T. Chaneli\`{e}re starting with a visit in Paris in Spring 2007, which triggered the investigations reported in this article. Furthermore, the authors thank Drs. R. Cone and C. Thiel for enlightening discussions on properties of Tm:LiNbO$_3$, in particular for their  groundbreaking work first reported at a workshop 2008 in Bozeman. We also thank  Dr. H. Suche for help regarding the waveguide characterization, and V. Kiselyov for technical support. Financial support by NSERC, GDC, iCORE, AET and CFI is acknowledged.

\section*{References}

\bibliographystyle{elsarticle-num}

\end{document}